\documentclass[twoside]{article}
\usepackage{qic}

\usepackage{amsmath}
\usepackage{amssymb}
\usepackage{braket}

\newtheorem{thm}{Theorem}

\newtheorem{lem}{Lemma}
\newtheorem{ex}{Example}

\begin{document}
\setlength{\textheight}{8.0truein}    

\runninghead{Limit theorems of a 3-state quantum walk and its application for discrete uniform measures}
            {T. Machida}

\normalsize\textlineskip
\thispagestyle{empty}
\setcounter{page}{1}


\vspace*{0.88truein}

\alphfootnote

\fpage{1}

\centerline{\bf
Limit theorems of a 3-state quantum walk and its application for discrete uniform measures}
\vspace*{0.37truein}
\centerline{\footnotesize
Takuya Machida}
\vspace*{0.015truein}
\centerline{\footnotesize\it Department of Mathematics, University of California,}
\baselineskip=10pt
\centerline{\footnotesize\it Berkeley, CA, 94720, USA}
\vspace*{0.015truein}
\centerline{\footnotesize\it Research Fellow of Japan Society for the Promotion of Science,}
\baselineskip=10pt
\centerline{\footnotesize\it Meiji University, Nakano Campus, 4-21-1 Nakano, Nakano-ku, Tokyo 164-8525, Japan}
\vspace*{0.225truein}

\vspace*{0.21truein}

\abstracts{
We present two long-time limit theorems of a 3-state quantum walk on the line when the walker starts from the origin.
One is a limit measure which is obtained from the probability distribution of the walk at a long-time limit, and the other is a convergence in distribution for the walker's position in a rescaled space by time.
In addition, as an application of the walk, we obtain discrete uniform limit measures from the 3-state walk with a delocalized initial state.  
}{}{}

\vspace*{10pt}


\vspace*{1pt}\textlineskip    

\bibliographystyle{qic}

\section{Introduction}

Investigation of quantum walks (QWs) started in early 2000's and various types of the quantum walks have been analyzed theoretically and numerically.
Although one can implement several experiments on QWs in physical systems, it is difficult to analyze statistically the QWs after a lot of steps.
Theoretical analysis plays an important role to understand the QWs after long time, and
long time limit theorems describe an asymptotic behavior of the QWs after many steps.
In this paper we treat a 3-state QW corresponding to a walk on the integers in which the walker moves to the left or the right, or stays at the current position at each time step according to a given probability law.
There are long time limit theorems for a 3-state walk on the integers.
Inui et al.~\cite{InuiKonnoSegawa2005} focused on a 3-state walk whose dynamics was determined by a Grover coin and derived two theorems.
One gives a limit measure of the walker's position, and the other is a convergence in distribution of a rescaled walker's position.
For the 3-state Grover walker starting from a certain position, localization can occur in the limit.
In this paper, localization will mean that the probability measure takes positive value at a certain position in the limit of infinite time. 
The limit measure exponentially decays with distance from the origin.
The probability distribution of the walker's position rescaled by time converges weakly to a random value which has both a Dirac $\delta$-function at the origin and a continuous function with a compact support.
The existence of the Dirac $\delta$-function denotes localization and it is caused by the eigenvalue 1 of a unitary operator which describes unit time evolution of the whole system of the 3-state Grover walk.  

Recently, it was proved in Machida~\cite{Machida2013} that a continuous uniform measure can be realized as a limit distribution by a 2-state QW with a delocalized initial state.
That is, the 2-state QW gives a quantum algorithm to create a continuous uniform measure approximately.
A uniform measure may be applied for building uniform random values.
To find an application of the QWs, it would be worth to study uniform measures from the view point of QWs.
So, we are interested in the relation between discrete measures and QWs.  
Our aim is to get discrete measures in a limit theorem from a 3-state QW, and the idea is based on localization of the walk and its limit measure.

In this paper we treat both a localized initial state at the origin and a delocalized initial state.
As mentioned above a 3-state Grover walk starting from the origin was analyzed in a limit by Inui et al.~\cite{InuiKonnoSegawa2005}.
On the other hand, there is no limit theorem for 3-state walks with a delocalized initial state.
Several models of 2-state QWs on the line with a delocalized initial state are numerically studied~\cite{AbalDonangeloRomanelliSiri2006,AbalSiriRomanelliDonangelo2006} and the possibility of uniform measures was predicted in Chandrashekar~\cite{Chandrashekar2010} and Valc\'{a}rcel et al.~\cite{ValcarcelRoldanRomanelli2010}.
In addition, some well-known probability distributions with a compact support including the Wigner semicircle, the Gaussian distribution, the arcsine distribution and so on, were realized by a 2-state QW in Machida~\cite{Machida2013,Machida2013c}.
The delocalized initial state of the 2-state walk is constructed of the Fourier coefficients of the functions we set, and we can create various limit distributions by controlling the functions.  
In the case of space-homogeneous 2-state walks, except for the obvious walks (e.g. the unit time-evolution operator is given by the unit matrix), we can not get discrete uniform measures because localization does not occur at all and the limit measure becomes zero at all positions.
Therefore we focus on 3-state walks, whose probability distribution can get localized, to realize discrete uniform measures.

In the subsequent sections, we will treat the definition of a 3-state walk, long-time limit theorems, their proofs by Fourier analysis, and an application of the walk for discrete uniform measures.
In Sec.~\ref{sec:definition} we define a 3-state QW on the line including a Grover walk.
Section~\ref{sec:limit_th} is devoted to the analysis of the 3-state walk starting from the origin.
As an application, we show that discrete uniform measures are realized in a 3-state walk with a delocalized state in Sec.~\ref{sec:application}.
In the final section, we conclude our results and discuss a future problem. 
    
\section{Definition of a discrete-time QW on the line}
\label{sec:definition}
Total system of discrete-time 3-state QWs on the line $\mathbb{Z}=\left\{0, \pm 1, \pm 2,\ldots\right\}$ is defined in a tensor space $\mathcal{H}_p\otimes\mathcal{H}_c$, where $\mathcal{H}_p$ is called a position Hilbert space which is spanned by a basis $\left\{\ket{x}:\,x\in\mathbb{Z}\right\}$ and $\mathcal{H}_c$ is called a coin Hilbert space which is spanned by a basis $\left\{\ket{0},\ket{1},\ket{2}\right\}$.
Let $\ket{\psi_{t}(x)} \in \mathcal{H}_c$ be the probability amplitudes of the walker at position $x$ at time $t \in\left\{0,1,2,\ldots\right\}$.
The state of the 3-state QWs on the line at time $t$ is expressed by $\ket{\Psi_t}=\sum_{x\in\mathbb{Z}}\ket{x}\otimes\ket{\psi_{t}(x)}\in\mathcal{H}_p\otimes\mathcal{H}_c$.
In this paper we deal with a 3-state walk whose coin-states are operated by
\begin{align}
 C=&-\frac{1+c}{2}\ket{0}\bra{0}+\frac{s}{\sqrt{2}}\ket{0}\bra{1}+\frac{1-c}{2}\ket{0}\bra{2}+\frac{s}{\sqrt{2}}\ket{1}\bra{0}+c\ket{1}\bra{1}+\frac{s}{\sqrt{2}}\ket{1}\bra{2}\nonumber\\
 &+\frac{1-c}{2}\ket{2}\bra{0}+\frac{s}{\sqrt{2}}\ket{2}\bra{1}-\frac{1+c}{2}\ket{2}\bra{2},
\end{align}
where $c=\cos\theta, s=\sin\theta$ with $\theta\in [0,2\pi)$.
The walks determined by $\theta=0,\pi$ are obvious.
So, we don't treat these cases.
We should note that when the parameter $\theta$ satisfies $\cos\theta=-\frac{1}{3}, \sin\theta=\frac{2\sqrt{2}}{3}$, the coin operator $C$ becomes a Grover coin.
The whole state at time $t+1$ is determined by $\ket{\Psi_{t+1}}=\tilde{S}\tilde{C}\ket{\Psi_t}$,
where
\begin{align}
 \tilde{C}=&\sum_{x\in\mathbb{Z}}\ket{x}\bra{x}\otimes C,\\
 \tilde{S}=&\sum_{x\in\mathbb{Z}}\ket{x-1}\bra{x}\otimes\ket{0}\bra{0}+\ket{x}\bra{x}\otimes\ket{1}\bra{1}+\ket{x+1}\bra{x}\otimes\ket{2}\bra{2}.
\end{align}
Assuming $\braket{\Psi_0|\Psi_0}=1$, we define the probability that the quantum walker with the state $\ket{j}\,(j=0,1,2)$ can be observed at position $x$ at time $t$ by
\begin{equation}
 \bra{\Psi_t}\left(\ket{x}\bra{x}\otimes\ket{j}\bra{j}\right)\ket{\Psi_t}.\label{eq:state_prob}
\end{equation}
The walker $X_t$, hence, can be found at position $x$ at time $t$ with probability
\begin{equation}
 \mathbb{P}(X_t=x)=\bra{\Psi_t}\left(\ket{x}\bra{x}\otimes\sum_{j=0}^2\ket{j}\bra{j}\right)\ket{\Psi_t}
\end{equation}
which is the sum of expression Eq.~(\ref{eq:state_prob}) over $j=0,1,2$.
We will concentrate on analyzing this probability $\mathbb{P}(X_t=x)$ as $t\to\infty$. 
While the definition of the time evolution has been given on the tensor Hilbert space $\mathcal{H}_p\otimes\mathcal{H}_c$, it is equivalently defined on the Fourier space 
\begin{equation}
\ket{\hat\Psi_t(k)}=\sum_{x\in\mathbb{Z}}e^{-ikx}\ket{\psi_t(x)}\quad(k\in\left[-\pi,\pi\right)).
\end{equation}
Then amplitudes are computed as the inverse Fourier transform
\begin{equation}
 \ket{\psi_t(x)}=\int_{-\pi}^\pi e^{ikx}\ket{\hat\Psi_t(k)}\frac{dk}{2\pi},
\end{equation}
and the time evolution of $\ket{\hat\Psi_t(k)}$ becomes
\begin{equation}
 \ket{\hat\Psi_{t+1}(k)}=\left(e^{ik}\ket{0}\bra{0}+\ket{1}\bra{1}+e^{-ik}\ket{2}\bra{2}\right)C\ket{\hat\Psi_t(k)}.
\end{equation}
The operator $e^{ik}\ket{0}\bra{0}+\ket{1}\bra{1}+e^{-ik}\ket{2}\bra{2}$ plays the same role as the position shift operator $\tilde{S}$.

\section{Limit theorem for the walk starting from the origin}
\label{sec:limit_th}

Long-time limit theorems of a 3-state QW on the line starting from the origin were firstly investigated in Inui et al.~\cite{InuiKonnoSegawa2005}. 
They derived a limit measure and a convergence theorem in distribution for a walk determined by a Grover coin.
{\v{S}}tefa{\v{n}}{\'a}k et al.~\cite{StefavnakBezdvekovaJex2012} got just peak velocities of a 3-state walk including the Grover walk.
The peak velocity plays an important role in giving the compact support of the limit distribution.
Recently, {\v{S}}tefa{\v{n}}{\'a}k et al.~\cite{vStefavnakBezdvekovaJexBarnett2014} analyzed a limit measure of the probability at the origin for a maximally mixed initial coin state which is given by stochastically mixing three initial coin states $\ket{\Psi_0}=\ket{0}\otimes\ket{0}, \ket{0}\otimes\ket{1}, \ket{0}\otimes\ket{2}$.
Let us take the initial state $\ket{\Psi_0}=\ket{0}\otimes\left(\alpha\ket{0}+\beta\ket{1}+\gamma\ket{2}\right)$ with $|\alpha|^2+|\beta|^2+|\gamma|^2=1$.
Then we obtain a limit measure for the 3-state QW determined by the coin operator $C$.
\begin{thm}
For $x\in\mathbb{Z}$, we have
 \begin{align}
  \lim_{t\to\infty}\mathbb{P}(X_t=x)=&\frac{1}{64(1-c)^2}\biggl\{2(1-c)\left|B\nu^{|x+1|}+A\nu^{|x|}\right|^2\nonumber\\
  &+(1+c)\left|B\nu^{|x+1|}+(A+B)\nu^{|x|}+A\nu^{|x-1|}\right|^2\nonumber\\
  &+2(1-c)\left|B\nu^{|x|}+A\nu^{|x-1|}\right|^2\biggr\},\label{eq:limit_measure_1}
 \end{align}
 where
 \begin{align}
  \nu=&\frac{-(3-c)+2\sqrt{2(1-c)}}{1+c}\,\in (-1,0),\\
  A=&2(1-c)\alpha+\sqrt{2}s\beta,\\
  B=&\sqrt{2}s\beta+2(1-c)\gamma.
 \end{align}
\label{th:limit1}
\end{thm}

Since we see
\begin{align}
 &\sum_{x\in\mathbb{Z}}\lim_{t\to\infty}\mathbb{P}(X_t=x)\nonumber\\
 =&\frac{1}{64(1-c)^2}\biggl[4(1-c)\biggl\{(|A|^2+|B|^2)\sum_{x\in\mathbb{Z}}\nu^{2|x|}+2\Re(A\overline{B})\sum_{x\in\mathbb{Z}}\nu^{|x|+|x+1|}\biggr\}\nonumber\\
 &+2(1+c)\biggl\{\Bigl(|A|^2+|B|^2+\Re(A\overline{B})\Bigr)\sum_{x\in\mathbb{Z}}\nu^{2|x|}
 +|A+B|^2\sum_{x\in\mathbb{Z}}\nu^{|x|+|x+1|}\nonumber\\
 &+\Re(A\overline{B})\sum_{x\in\mathbb{Z}}\nu^{|x+1|+|x-1|}\biggr\}\biggr]\nonumber\\
 =&\frac{1}{8\sqrt{2}(1-c)^{\frac{3}{2}}}\left\{|A|^2+|B|^2+2\nu\Re(A\overline{B})\right\},\label{eq:sum_limP}
\end{align}
the sum over all integers takes a value less than 1 depending on the parameters $\theta,\alpha,\beta,\gamma$, where $\Re(z)$ denotes the real part of the complex number $z$.
Here, we define localization of the walk by $\exists\,x\in\mathbb{Z},\,\,\lim_{t\to\infty}\mathbb{P}(X_t=x)>0$, and the sum leads us to the condition that the walker localizes:
\begin{equation}
 |A|^2+|B|^2+2\nu\Re(A\overline{B})>0.
\end{equation}

\begin{figure}[h]
\begin{center}
 \begin{minipage}{60mm}
  \begin{center}
  \includegraphics[scale=0.5]{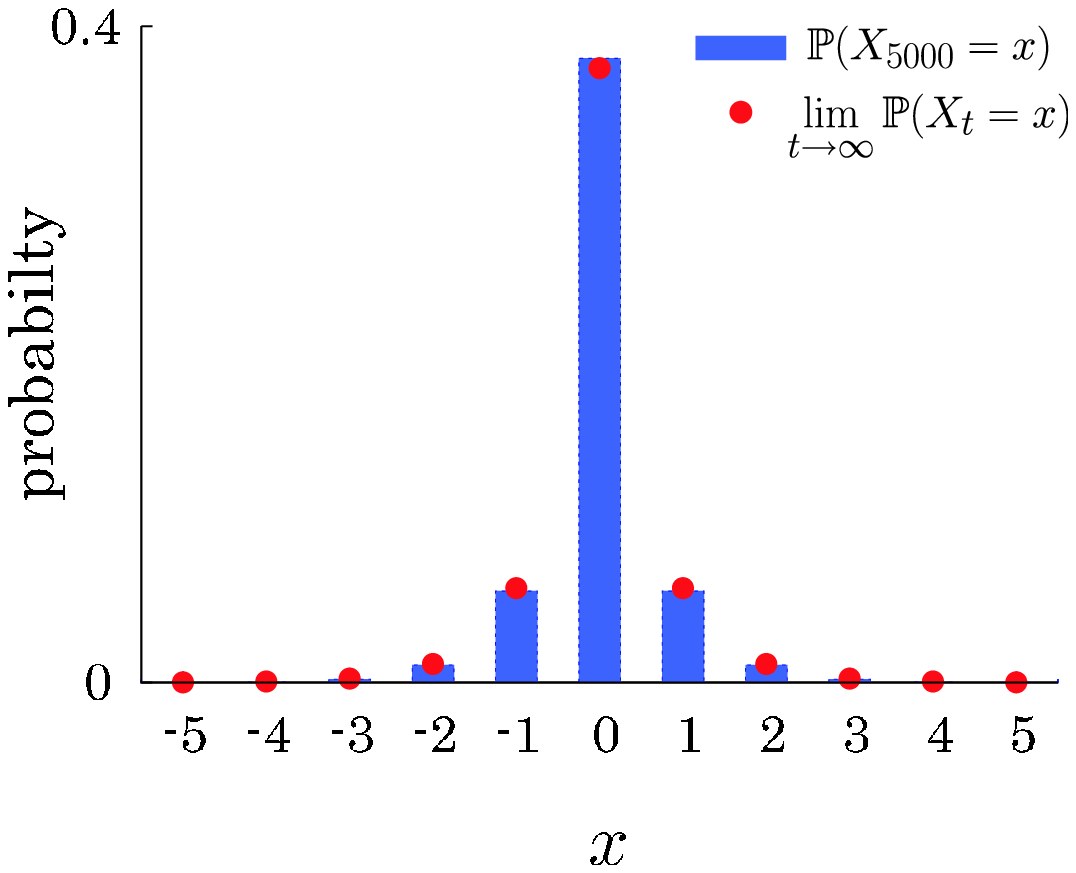}\\[2mm]
  (a)
  \end{center}
 \end{minipage}
 \begin{minipage}{60mm}
  \begin{center}
 \includegraphics[scale=0.5]{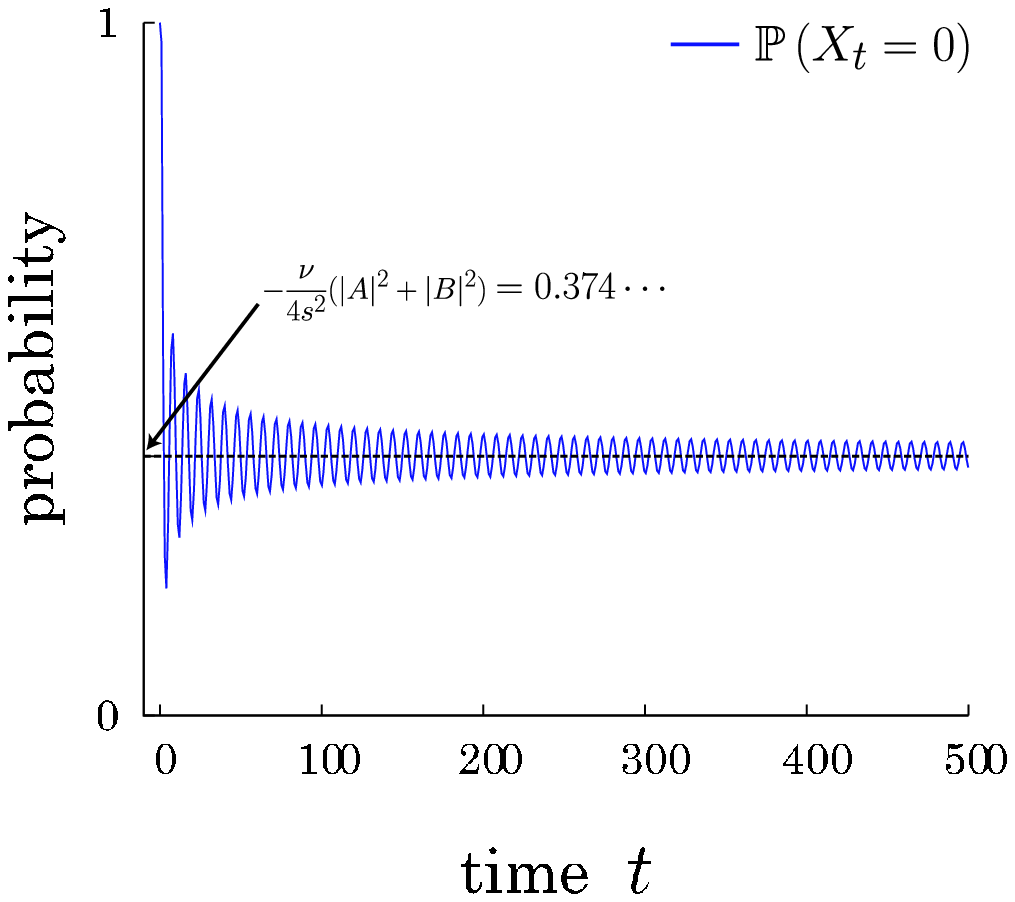}\\[2mm]
  (b)
  \end{center}
 \end{minipage}
 \vspace{5mm}
 \fcaption{We take $\theta=\frac{\pi}{4},\, \alpha=\beta=\gamma=\frac{1}{\sqrt{3}}$ in Figs.~(a) and (b). Figure~(a) shows a comparison between probability distribution at time $t=5000$ (blue bars) and the limit measure $\lim_{t\to\infty}\mathbb{P}(X_t=x)$ (red points). We can see the convergence of the probability at the origin (blue line) for time $t$ in Fig.~(b).}
 \label{fig:time_vs_limit}
\end{center}
\end{figure}

\begin{figure}[h]
\begin{center}
 \begin{minipage}{90mm} 
  \begin{center}
   \includegraphics[scale=0.5]{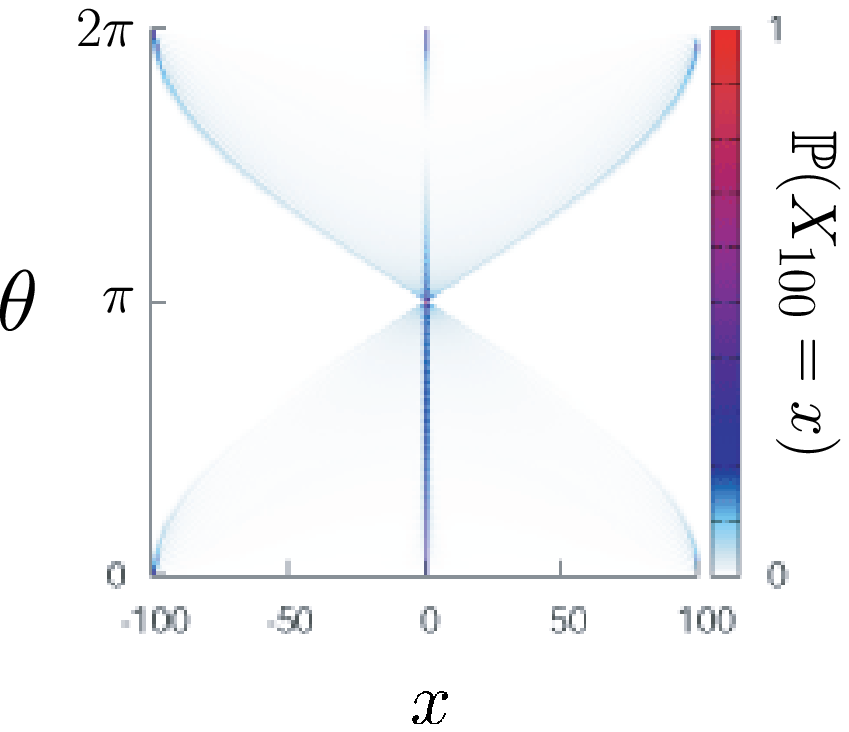}
   \includegraphics[scale=0.5]{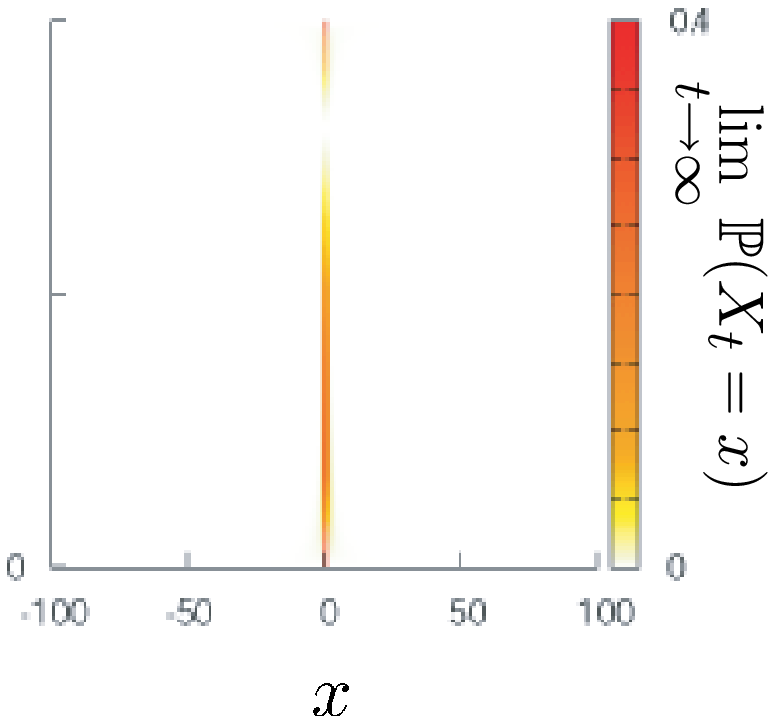}\\[2mm]
   (a)
  \end{center}
 \end{minipage}
 \begin{minipage}{45mm}
  \begin{center}
   \includegraphics[scale=0.4]{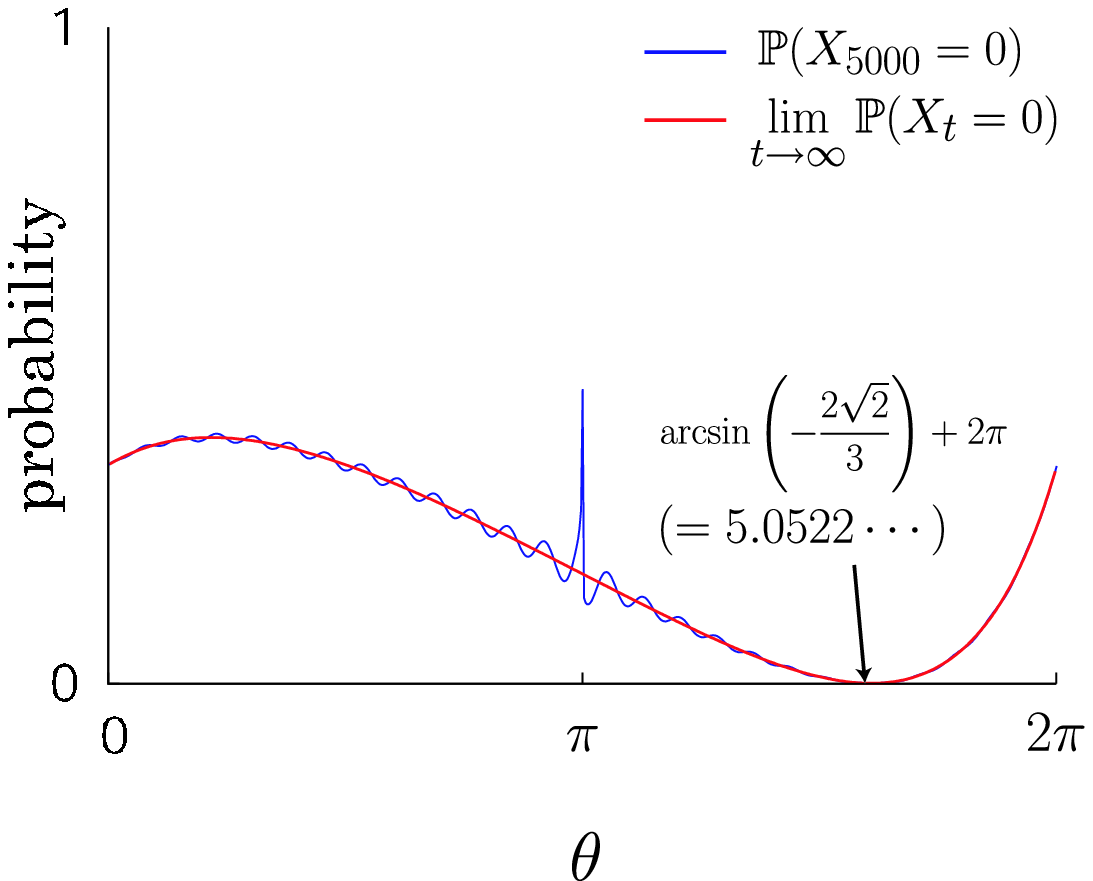}\\[2mm]
   (b)
  \end{center}
 \end{minipage}
 \vspace{5mm}
 \fcaption{Figures~(a) and (b) show probability distributions at a certain time and the corresponding limit measure for the parameter $\theta$ of the coin operator $C$ in the case of  $\alpha=\beta=\gamma=\frac{1}{\sqrt{3}}$. The left side in Fig.~(a) is probability distributions at time $t=100$. Regarding the origin, Fig.~(b) lets us know a discrepancy between the probability at time $t=5000$ (blue line) and the limit measure (red line).}
 \label{fig:theta_vs_limit}
\end{center}
\end{figure}

Figures~\ref{fig:time_vs_limit} and \ref{fig:theta_vs_limit} support Theorem~\ref{th:limit1} as a comparison between the limit measure $\lim_{t\to\infty}\mathbb{P}(X_t=x)$ and the probability distribution $\mathbb{P}(X_t=x)$ at an infinite time.
The initial state at the origin is set as $\alpha=\beta=\gamma=\frac{1}{\sqrt{3}}$ in each figure.
Since $\theta=\frac{\pi}{4}$ in Fig.~\ref{fig:time_vs_limit}, the limit value of the probability at the origin becomes $\lim_{t\to\infty}\mathbb{P}(X_t=0)=-\frac{\nu}{4s^2}(|A|^2+|B|^2)=0.374\cdots$.
Also, we can visually confirm in Fig.~\ref{fig:theta_vs_limit}-(b) that the limit measure $\lim_{t\to\infty}\mathbb{P}(X_t=0)$ is zero when we take $\theta=\arcsin\left(-\frac{2\sqrt{2}}{3}\right)+2\pi=5.0522\cdots$, for $\alpha=\beta=\gamma=\frac{1}{\sqrt{3}}$.
Those parameters actually produce $A=B=0$.
Also, we see that the walk repeated by them does not localize in our definition of localization.

\begin{proof}{
We are mainly making a calculation on the Fourier space in our proof.
The Fourier method for discrete-time quantum walks was introduced in Grimmett et al.~\cite{GrimmettJansonScudo2004}. 

At the beginning, to analyze the walk, we put $\ket{0}={}^T[1,0,0],\,\ket{1}={}^T[0,1,0],\,\ket{2}={}^T[0,0,1]$, where $T$ is the transposed operator.
Then the operator $\hat{C}(k)=\left(e^{ik}\ket{0}\bra{0}+\ket{1}\bra{1}+e^{-ik}\ket{2}\bra{2}\right)C$ for the Fourier transform $\ket{\hat\Psi_t(k)}$ is rewritten as
\begin{equation}
 \hat{C}(k)=\left[\begin{array}{ccc}
	     e^{ik}& 0& 0\\
	     0& 1& 0\\
	     0& 0& e^{-ik}
		  \end{array}\right]
 \left[\begin{array}{ccc}
  -\frac{1+c}{2} & \frac{s}{\sqrt{2}} & \frac{1-c}{2}\\
	\frac{s}{\sqrt{2}} & c & \frac{s}{\sqrt{2}}\\
	\frac{1-c}{2} & \frac{s}{\sqrt{2}} & -\frac{1+c}{2}
       \end{array}\right],
\end{equation} 
and it has the eigenvalues
\begin{align}
 \lambda_1(k)=&1,\\
 \lambda_j(k)=&\frac{-\left\{(1+c)\cos k+(1-c)\right\}+i(-1)^j\sqrt{4-\left\{(1+c)\cos k+1-c\right\}^2}}{2}\quad(j=2,3).
\end{align}
Here, we realize that the coin operator $C$ is a product of five rotation matrices
\begin{align}
  & \left[\begin{array}{ccc}
  -\frac{1+c}{2} & \frac{s}{\sqrt{2}} & \frac{1-c}{2}\\
	\frac{s}{\sqrt{2}} & c & \frac{s}{\sqrt{2}}\\
	\frac{1-c}{2} & \frac{s}{\sqrt{2}} & -\frac{1+c}{2}
       \end{array}\right]\nonumber \\
  =&\left[\begin{array}{ccc}
    1& 0 &0 \\
	  0& 0 & -1\\
	  0 & 1& 0
	 \end{array}\right]
  \left[\begin{array}{ccc}
   \frac{1}{\sqrt{2}} & -\frac{1}{\sqrt{2}} & 0\\
	 \frac{1}{\sqrt{2}} & \frac{1}{\sqrt{2}} & 0\\
	 0 & 0 & 1
	\end{array}\right]
  \left[\begin{array}{ccc}
   -c & 0 & s\\
	 0 & 1 & 0\\
	 -s & 0 & -c
	\end{array}\right]
  \left[\begin{array}{ccc}
   \frac{1}{\sqrt{2}} & -\frac{1}{\sqrt{2}} & 0\\
	 \frac{1}{\sqrt{2}} & \frac{1}{\sqrt{2}} & 0\\
	 0 & 0 & 1
	\end{array}\right]
  \left[\begin{array}{ccc}
    1& 0 &0 \\
	  0& 0 & -1\\
	  0 & 1& 0
	 \end{array}\right].
\end{align}
One of the forms of the eigenvectors $\ket{w_j(k)}\,(j=1,2,3)$ corresponding to $\lambda_j(k)$ becomes
\begin{equation}
 \ket{w_j(k)}=\left[\begin{array}{c}
	       \frac{1}{1+\lambda_j(k)e^{-ik}}\\[2mm]
		     \frac{\sqrt{2}s}{(1-c)(1+\lambda_j(k))}\\[2mm]
		     \frac{1}{1+\lambda_j(k)e^{ik}}
		    \end{array}\right].
\end{equation}
Decomposing the Fourier transform at time $t=0$ to three orthonormal eigenvectors, we have
\begin{align}
 \ket{\hat\Psi_t(k)}=&\hat{C}(k)^t \ket{\hat\Psi_0(k)}=\hat{C}(k)^t \sum_{j=1}^3 \braket{v_j(k)|\hat\Psi_0(k)}\ket{v_j(k)}\nonumber\\
 =&\sum_{j=1}^3 \lambda_j(k)^t \braket{v_j(k)|\hat\Psi_0(k)}\ket{v_j(k)},
\end{align}
where $\ket{v_j(k)}=\ket{w_j(k)}/\braket{w_j(k)|w_j(k)}$.
The Riemann-Lebesgue lemma derives the asymptotic behavior of probability amplitudes
\begin{align}
 \ket{\psi_t(x)}=&\sum_{j=1}^3 \int_{-\pi}^\pi e^{ikx}\lambda_j(k)^t \braket{v_j(k)|\hat\Psi_0(k)}\ket{v_j(k)}\,\frac{dk}{2\pi}\nonumber\\
 \sim&\int_{-\pi}^\pi e^{ikx}\braket{v_1(k)|\hat\Psi_0(k)}\ket{v_1(k)}\,\frac{dk}{2\pi}\quad (t\to\infty)\nonumber\\
 =&\frac{1+c}{8s^2\sqrt{2(1-c)}}\left[\begin{array}{c}
				 2(1-c)\left(B\nu^{|x+1|}+A\nu^{|x|}\right)\\[2mm]
				       \sqrt{2}s\left\{B\nu^{|x+1|}+(A+B)\nu^{|x|}+A\nu^{|x-1|}\right\}\\[2mm]
				       2(1-c)\left(B\nu^{|x|}+A\nu^{|x-1|}\right)
				      \end{array}\right],\label{eq:approximate}
\end{align}
where $g(t)\sim h(t)\,(t\to\infty)$ means $\lim_{t\to\infty}g(t)/h(t)=1$ and $\nu=\frac{-(3-c)+2\sqrt{2(1-c)}}{1+c},\, A=2(1-c)\alpha+\sqrt{2}s\beta,\, B=\sqrt{2}s\beta+2(1-c)\gamma$.
One can reach the statement of Theorem 1 from Eq.~(\ref{eq:approximate}).
}
\end{proof}

Theorem~\ref{th:limit1} describes probability distribution of $X_t$ as $t\to\infty$.
Next we present a limit distribution of the spatially rescaled random value $X_t/t$.  
It also expresses an asymptotic behavior of the walk after long time.   
The limit distribution is constructed of the Dirac $\delta$-function at the origin and a continuous function with a compact support which depends on the coin operator $C$.
We will be able to see that a bilateral relationship is linked by the coefficient of the Dirac $\delta$-function.

\begin{thm}
For the real number $x$, we have
 \begin{equation}
   \lim_{t\to\infty}\mathbb{P}\left(\frac{X_t}{t}\leq x\right)=\int_{-\infty}^x \Delta\delta_0(y)+f(y)I_{\left(-\sqrt{\frac{1+c}{2}},\sqrt{\frac{1+c}{2}}\right)}(y)\,dy,
 \end{equation}
 where $\delta_0(x)$ is the Dirac $\delta$-function at the origin and
 \begin{align}
  \Delta=&\frac{1}{8\sqrt{2}(1-c)^{\frac{3}{2}}}\left\{|A|^2+|B|^2+2\nu\Re(A\overline{B})\right\},\\
  f(x)=&\frac{\sqrt{1-c}}{2\pi(1-x^2)\sqrt{1+c-2x^2}}(d_0+d_1x+d_2x^2),\\
  d_0=&|\alpha+\gamma|^2+2|\beta|^2,\\
  d_1=&2\left\{-|\alpha-\beta|^2+|\gamma-\beta|^2-\left(2-\frac{\sqrt{2}s}{1+c}\right)\Re\left((\alpha-\gamma)\overline{\beta}\right)\right\},\\
  d_2=&|\alpha|^2-2|\beta|^2+|\gamma|^2-2\left\{\frac{\sqrt{2}s}{1+c}\Re((\alpha+\gamma)\overline{\beta})+\frac{3-c}{1+c}\Re(\alpha\overline{\gamma})\right\},\\
  I_\Gamma(x)=&\left\{\begin{array}{cl}
	   1&(x\in \Gamma),\\
		  0&(x\notin \Gamma).
		 \end{array}\right.
 \end{align}
 \label{th:limit2}
\end{thm}

The limit density function of the 3-state walk spatially rescaled by time has a compact support and its domain $\left(-\sqrt{\frac{1+c}{2}},\sqrt{\frac{1+c}{2}}\,\right)$ is perfectly dictated by the $\ket{0}\bra{0}$-component of the coin operator $C$.
This fact agrees with a result in {\v{S}}tefa{\v{n}}{\'a}k et al.~\cite{StefavnakBezdvekovaJex2012} (See Eq.~(16) in their paper), and a parameter $\rho$ in Eq.~(14) in their paper corresponds to $\sqrt{\frac{1+c}{2}}$. 
Parameters $c=-\frac{1}{3}, s=\frac{2\sqrt{2}}{3}$, moreover, produces the same limit distribution derived in Inui et al.~\cite{InuiKonnoSegawa2005}.

\begin{figure}[h]
\begin{center}
 \begin{minipage}{70mm}
  \begin{center}
  \includegraphics[scale=0.5]{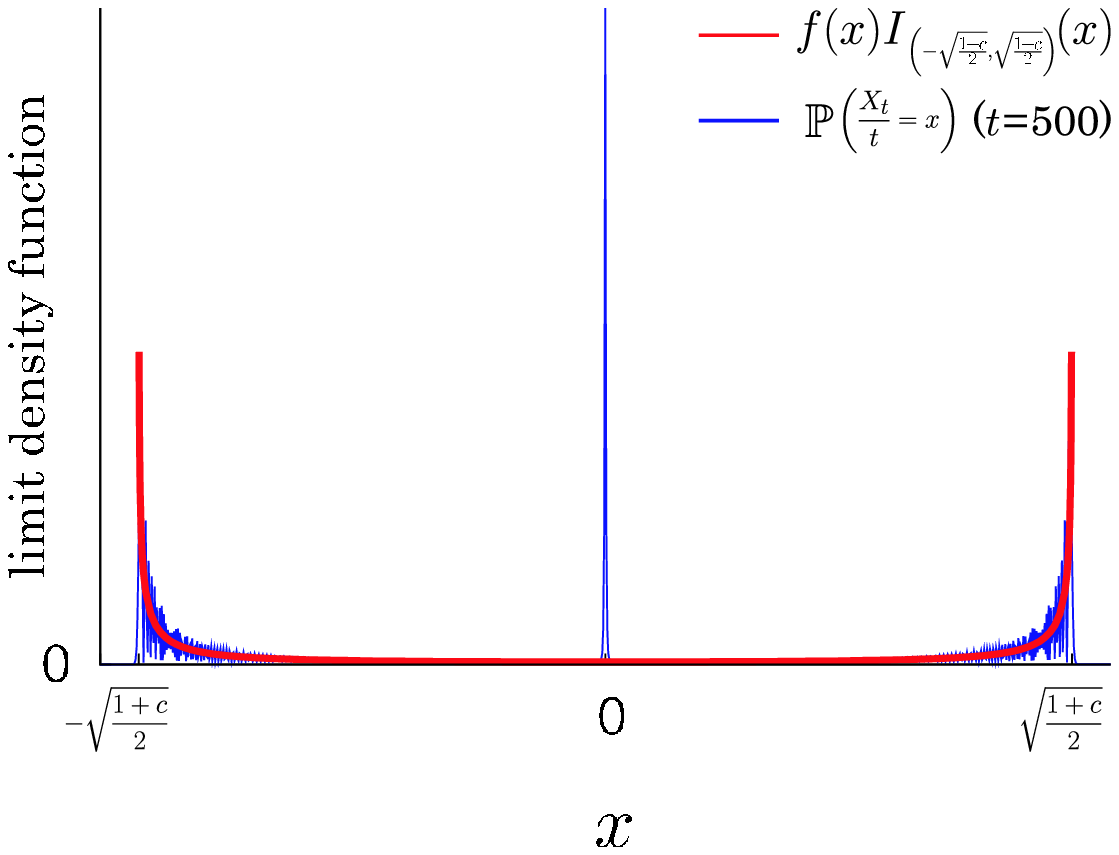}\\[2mm]
  (a) $\alpha=\beta=\gamma=\frac{1}{\sqrt{3}}$
  \end{center}
 \end{minipage}
 \begin{minipage}{70mm}
  \begin{center}
   \includegraphics[scale=0.5]{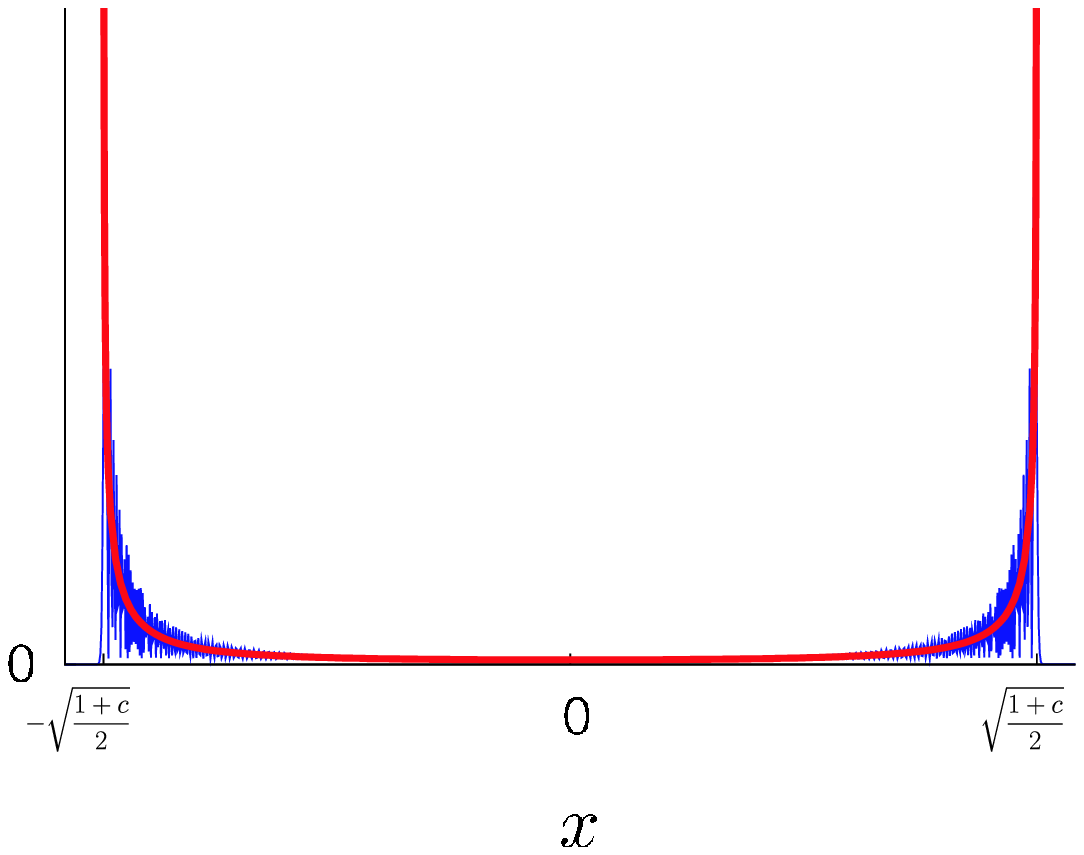}\\[2mm]
  (b) $\alpha=\gamma=\frac{1}{2\sqrt{2-\sqrt{2}}}, \beta=-\frac{\sqrt{2-\sqrt{2}}}{2}$
  \end{center}
 \end{minipage}
\vspace{5mm}
\fcaption{These figures show a comparison between probability distribution at time $t=500$ and the continuous part of the limit density function in the case of $\theta=\frac{\pi}{4}$.}
\label{fig:limit2}
\end{center}
\end{figure}

When we take $\theta=\frac{\pi}{4}$, we can see the continuous part of the limit density function $f(x)I_{\left(-\sqrt{\frac{1+c}{2}},\sqrt{\frac{1+c}{2}}\right)}(x)$ and the rescaled probability distribution $\mathbb{P}\left(\frac{X_t}{t}=x\right)$ at time $t=500$ in Fig.~\ref{fig:limit2}.
For $\theta=\frac{\pi}{4}$, the initial state with $\alpha=\gamma=\frac{1}{2\sqrt{2-\sqrt{2}}}, \beta=-\frac{\sqrt{2-\sqrt{2}}}{2}$ holds $\Delta=0$.
That means the probability distribution at time $t=500$ does not have a large mass around the origin in Fig.~\ref{fig:limit2}-(b), differently from Fig.~\ref{fig:limit2}-(a).

\begin{proof}{
What we want to show is convergence of the $r$-th moment $\mathbb{E}\left[(X_t/t)^r\right]$ ($r=0,1,2,\ldots$), because it is equivalent to getting the generating function $\mathbb{E}[e^{izX_t/t}]$. 

For $r=0,1,2,\ldots$,
\begin{align}
 \mathbb{E}(X_t^r)=&\sum_{x\in\mathbb{Z}}x^r\mathbb{P}(X_t=x)\nonumber\\
 =&\int_{-\pi}^\pi \bra{\hat\Psi_t(k)}\left(D^r\ket{\hat\Psi_t(k)}\right)\frac{dk}{2\pi}\nonumber\\
 =&(t)_r\int_{-\pi}^\pi \sum_{j=1}^3 \left(\frac{i\lambda'_j(k)}{\lambda_j(k)}\right)^r\left|\braket{v_j(k)|\hat\Psi_0(k)}\right|^2\frac{dk}{2\pi}+O(t^{r-1}),
 \label{eq:r-th_moment}
\end{align}
where $D=i(d/dk)$ and $(t)_r=t(t-1)\times\cdots\times(t-r+1)$.
We divide both sides of Eq.~(\ref{eq:r-th_moment}) by $t^r$ and take a limit 
\begin{align}
 \lim_{t\to\infty}\mathbb{E}\left[\left(\frac{X_t}{t}\right)^r\right]=&\int_{-\pi}^\pi 0^r \left|\braket{v_1(k)|\hat\Psi_0(k)}\right|^2\frac{dk}{2\pi}+\sum_{j=2}^3 \int_{-\pi}^\pi \left(\frac{i\lambda'_j(k)}{\lambda_j(k)}\right)^r\left|\braket{v_j(k)|\hat\Psi_0(k)}\right|^2\frac{dk}{2\pi},
\end{align}
where
\begin{equation}
 \frac{i\lambda'_j(k)}{\lambda_j(k)}=(-1)^j\frac{(1+c)\sin k}{\sqrt{4-\left\{(1+c)\cos k+(1-c)\right\}^2}}\quad (j=2,3).
\end{equation}
After making a direct calculation of the first term and putting $i\lambda'_j(k)/\lambda_j(k)=x$ in the other integrals, we get
\begin{align}
 \lim_{t\to\infty}\mathbb{E}\left[\left(\frac{X_t}{t}\right)^r\right]=&0^r\Delta+\int_{-\sqrt{\frac{1+c}{2}}}^{\sqrt{\frac{1+c}{2}}} \,x^r f(x)\,dx\nonumber\\
 =&\int_{-\infty}^\infty x^r \left\{\Delta\delta_0(x)+f(x)I_{\left(-\sqrt{\frac{1+c}{2}},\sqrt{\frac{1+c}{2}}\right)}(x)\right\}\,dx.
 \label{eq:r-th_moment2}
\end{align}
Equation~(\ref{eq:r-th_moment2}) means that $X_t/t$ converges in distribution to a random value which has the density $\Delta\delta_0(x)+f(x)I_{\left(-\sqrt{\frac{1+c}{2}},\sqrt{\frac{1+c}{2}}\right)}(x)\,dx$.
}
\end{proof}

\section{Application for discrete uniform measures}
\label{sec:application}

Throughout this section we prove that a 3-state QW with a delocalized initial state can create uniform measures.
Realization of uniform measures is inextricably linked to that of uniform random values, and it is a research subject in computer science.
Uniform measures are useful to generate various kind of distributions by a technical method (e.g. the inverse transform sampling method~\cite{Devroye1986}, the Box--Muller transform~\cite{BoxMuller1958}).
It have been already verified in Machida~\cite{Machida2013} that a continuous uniform probability measure can be constructed of a 2-state QW on the line starting from a delocalized initial state.

Given a function $F : \mathbb{Z} \longmapsto \mathbb{C}$ which satisfies
\begin{align}
 \sum_{x\in\mathbb{Z}}\left|\frac{1}{2}F(x-1)+\frac{3-c}{1+c}F(x)+\frac{1}{2}F(x+1)\right|^2=1,\\
 \forall\,y\in\mathbb{Z},\quad\int_{-\pi}^\pi\sum_{x\in\mathbb{Z}} e^{-ik(x-y)}F(x)\,dk=\sum_{x\in\mathbb{Z}}\int_{-\pi}^\pi e^{-ik(x-y)}F(x)\,dk,\label{eq:condition_F_2}
\end{align}
we take an initial state
\begin{equation}
 \ket{\psi_0(x)}=\left\{\frac{1}{2}F(x-1)+\frac{3-c}{1+c}F(x)+\frac{1}{2}F(x+1)\right\}\ket{\phi},\label{eq:delocalized_ini}
\end{equation}
where $\ket{\phi}=\alpha\ket{0}+\beta\ket{1}+\gamma\ket{2}$ such that $|\alpha|^2+|\beta|^2+|\gamma|^2=1$.
Applying the method to derive Theorem~\ref{th:limit1} for the 3-state walk with Eq.~(\ref{eq:delocalized_ini}), we obtain a limit measure
 \begin{align}
   \lim_{t\to\infty}\mathbb{P}(X_t=x)=&\frac{1}{4(1+c)}\biggl\{\frac{1}{1+c}\left|AF(x)+BF(x+1)\right|^2\nonumber\\
  &\frac{1}{2(1-c)}\left|AF(x-1)+(A+B)F(x)+BF(x+1)\right|^2\nonumber\\
  &\frac{1}{1+c}\left|AF(x-1)+BF(x)\right|^2\biggr\}.\label{eq:limit_measure_2}
 \end{align}
We utilize this limit measure for realization of a discrete uniform measure.
Before designing a uniform measure by the 3-state QW, we introduce a lemma for a special function $F(x)$. 

\begin{lem}
 For a positive integer $n$, we set
 \begin{equation}
  M(n)=\left\{1+\left(\frac{3-c}{1+c}\right)^2\right\}n-\frac{1}{2}.
 \end{equation}
 Then a function $F(x)$ such as
 \begin{equation}
  F(x)=\left\{\begin{array}{cl}
	\frac{1}{\sqrt{M(n)}}&(x=0,2,4,\ldots,2n-2),\\
	       0&(\mbox{otherwise}),
	      \end{array}\right.
 \end{equation}
 produces a delocalized initial state 
 \begin{equation}
  \ket{\psi_0(x)}=\left\{\begin{array}{cl}
		   \frac{1}{2\sqrt{M(n)}}\ket{\phi}& (x=-1,2n-1),\\[3mm]
		   \frac{3-c}{(1+c)\sqrt{M(n)}}\ket{\phi}& (x=0,2,\ldots,2n-2),\\[3mm]
		   \frac{1}{\sqrt{M(n)}}\ket{\phi}& (x=1,3,\ldots,2n-3),\\[3mm]
		   0\ket{\phi}& (\mbox{otherwise}),
			 \end{array}\right.
 \end{equation}
 and a limit measure
 \begin{equation}
  \lim_{t\to\infty}\mathbb{P}(X_t=x)=\left\{\begin{array}{cl}
				      \frac{(3-c)|B|^2}{8(1+c)s^2M(n)}&(x=-1),\\[3mm]
					     \frac{1}{4(1+c)M(n)}\left\{\frac{|A|^2+|B|^2}{1+c}+\frac{|A+B|^2}{2(1-c)}\right\}&(x=0,1,\ldots,2n-2),\\[3mm]
					      \frac{(3-c)|A|^2}{8(1+c)s^2M(n)}&(x=2n-1),\\[3mm]
					      0&(\mbox{otherwise}).
					    \end{array}\right.
 \end{equation}
\label{lemma}
\end{lem}
For the function $F(x)$ treated in Lemma~\ref{lemma}, the sum $\sum_{x\in\mathbb{Z}} e^{-ik(x-y)}F(x)$ is reduced to a finite sum over $x=0,2,4,\ldots,2n-2$.
One can interchange the summation with the integration in Eq.~(\ref{eq:condition_F_2}) clearly.

As we can see in the below examples, Lemma~\ref{lemma} presents discrete uniform measures.

\begin{ex}

When we take $A=0, B\neq 0$ (e.g. $\alpha=\beta=0, \gamma=1$), a $2n$-point discrete uniform measure is obtained.
\begin{equation}
 \lim_{t\to\infty}\mathbb{P}(X_t=x)=\left\{\begin{array}{cl}
				     \frac{(3-c)|B|^2}{8(1+c)s^2M(n)}&(x=-1,0,1,\ldots,2n-2),\\
					    0&(\mbox{otherwise}).
					   \end{array}\right.
\end{equation}
\label{ex:1}
\end{ex}

\begin{ex}

When we take $A\neq 0, B=0$ (e.g. $\alpha=1, \beta=\gamma=0$), a $2n$-point discrete uniform measure is obtained.
\begin{equation}
 \lim_{t\to\infty}\mathbb{P}(X_t=x)=\left\{\begin{array}{cl}
				     \frac{(3-c)|A|^2}{8(1+c)s^2M(n)}&(x=0,1,\ldots,2n-1),\\
					    0&(\mbox{otherwise}).
					   \end{array}\right.
\end{equation}
\label{ex:2}
\end{ex}

\begin{ex}
When we take $B=-A (\neq 0)$,
\begin{equation}
 \lim_{t\to\infty}\mathbb{P}(X_t=x)=\left\{\begin{array}{cl}
				     \frac{(3-c)|A|^2}{8(1+c)s^2M(n)}&(x=-1,2n-1),\\
					    \frac{|A|^2}{2(1+c)^2M(n)}&(x=0,1,\ldots,2n-2),\\
					    0&(\mbox{otherwise}).
					   \end{array}\right.
\end{equation}
Particularly setting $c=\frac{1}{3}, s=\pm\frac{2\sqrt{2}}{3}, \beta=\mp\frac{1}{2}(\alpha+\gamma)$ with $\gamma\neq \alpha$ respectively, we see a $2n+1$-point discrete uniform measure
\begin{equation}
 \lim_{t\to\infty}\mathbb{P}(X_t=x)=\left\{\begin{array}{cl}
				     \frac{|\alpha-\gamma|^2}{4(10n-1)}&(x=-1,0,1,\ldots,2n-1),\\
					    0&(\mbox{otherwise}).
					   \end{array}\right.
\end{equation}
We should note that the condition $B=-A (\neq 0)$ is equivalent to $\beta=\mp\frac{1}{2}(\alpha+\gamma)$ with $\gamma\neq \alpha$ when $c=\frac{1}{3}, s=\pm\frac{2\sqrt{2}}{3}$, respectively.
\label{ex:3}
\end{ex}

\begin{figure}[h]
\begin{center}
 \begin{minipage}{50mm}
  \begin{center}
  \includegraphics[scale=0.4]{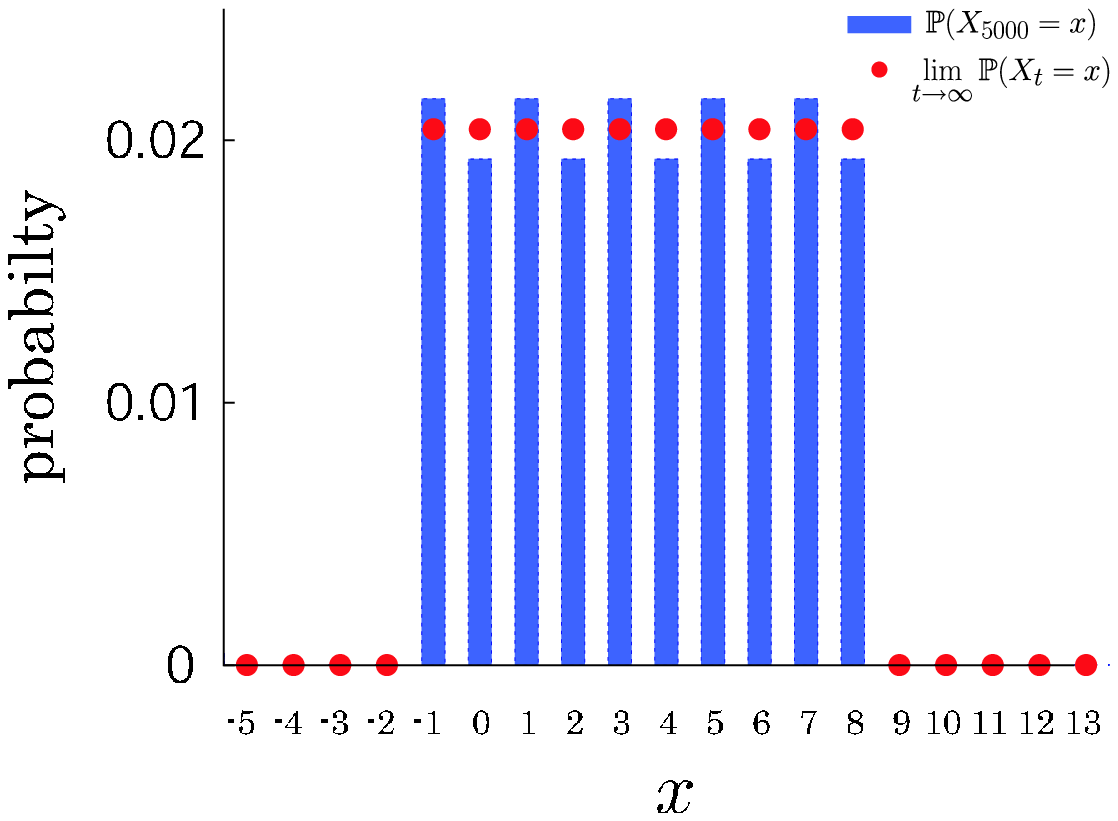}\\[2mm]
  (a) Example~\ref{ex:1}\\ $\alpha=\beta=0, \gamma=1$
  \end{center}
 \end{minipage}
 \begin{minipage}{40mm}
  \begin{center}
   \includegraphics[scale=0.4]{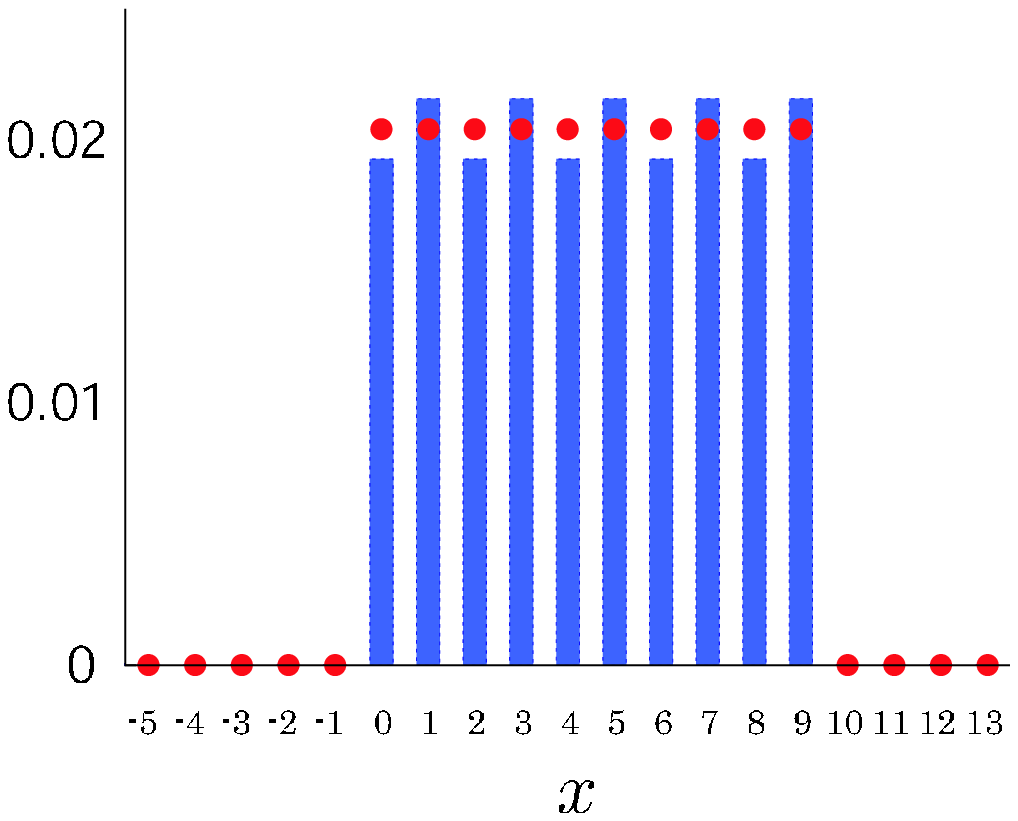}\\[2mm]
  (b) Example~\ref{ex:2}\\ $\alpha=1, \beta=\gamma=0$
  \end{center}
 \end{minipage}
 \begin{minipage}{40mm}
  \begin{center}
   \includegraphics[scale=0.4]{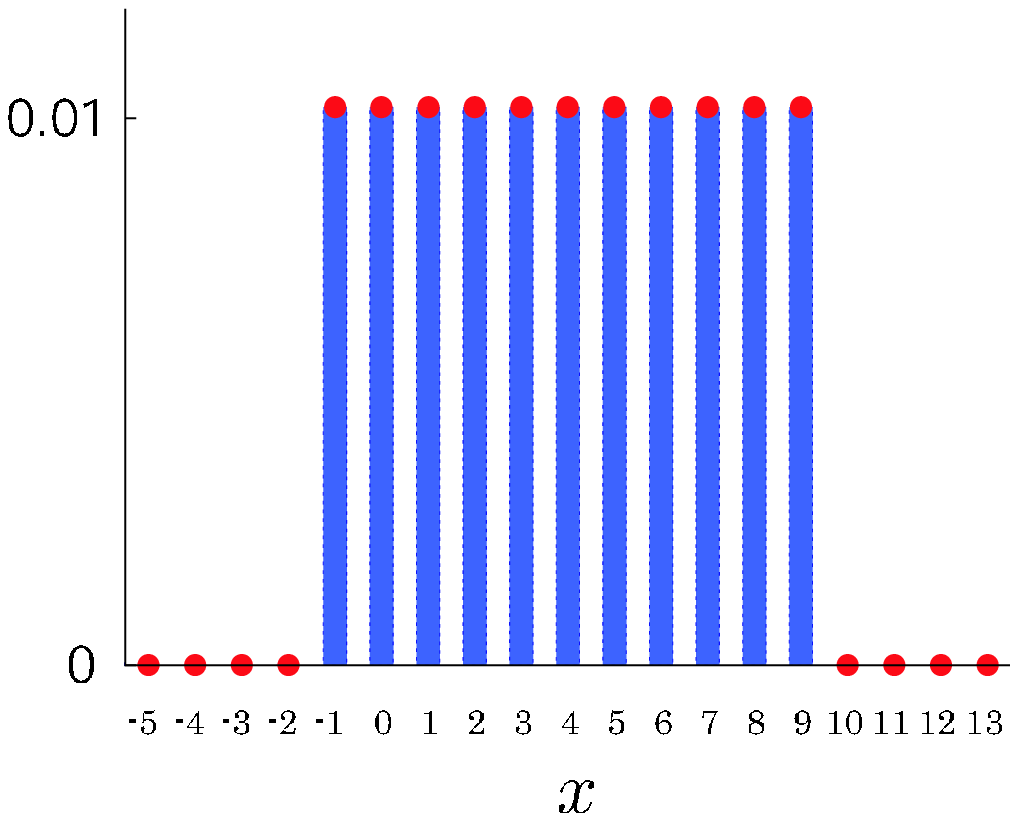}\\[2mm]
  (c) Example~\ref{ex:3}\\ $\alpha=-\gamma=\frac{1}{\sqrt{2}},\beta=0$
  \end{center}
 \end{minipage}
\end{center}
\vspace{5mm}
\fcaption{As $c=\frac{1}{3},s=\frac{2\sqrt{2}}{3},n=5$, Examples~\ref{ex:1}, \ref{ex:2} and \ref{ex:3} produce Figs. (a), (b) and (c). Blue bars and red points mean probability distribution $\mathbb{P}(X_{5000}=x)$ and the limit measure $\lim_{t\to\infty}\mathbb{P}(X_t=x)$, respectively.}
\label{fig:ex123}
\end{figure}

Figure~\ref{fig:ex123} broadens our understanding of the relationship between probability distribution at time $t=5000$ and the limit measure given in Examples~\ref{ex:1}, \ref{ex:2} and \ref{ex:3}.
The parameters are set as $c=\frac{1}{3}, s=\frac{2\sqrt{2}}{3}, n=5$ in Figs.~\ref{fig:ex123}-(a), -(b) and -(c).
In Figs.~~\ref{fig:ex123}-(a) and -(b) (resp. Fig~~\ref{fig:ex123}-(c)), a $10$-point (resp. 11-point) discrete uniform measure is realized as $t\to\infty$.

\section{Summary and Discussion}
\label{sec:summary}

In this paper we presented two limit theorems for a 3-state QW starting from the origin and proved to construct a uniform measure by using the 3-state walk with a delocalized initial state.
When the walker starts from the origin, the limit measure exponentially decayed from the origin.
We should note that it was not probability measure.
The sum of the limit values over all positions, however, was informative on localization of probability distribution.
To know whether the walker localizes or not, we have only to consider the quantity $|A|^2+|B|^2+2\nu\Re(A\overline{B})$.
If we give an initial state $\alpha=\gamma=0,\beta=1$ to the 3-state walk starting from the origin, the walker localizes because of $|A|^2+|B|^2+2\nu\Re(A\overline{B})=4s^2(1+\nu)>0$.
Considering the walker operated by the coin operator with $\theta=0,\pi$ localizes, we realize localization is occurred as long as we set $\alpha=\gamma=0,\beta=1$.
Whereas, if we take an initial state $\alpha=\beta=\gamma=\frac{1}{\sqrt{3}}$, the walker localizes for all but the case $c=\frac{1}{3},s=-\frac{2\sqrt{2}}{3}$ (See Fig~\ref{fig:theta_vs_limit}-(b)).

In Theorem~\ref{th:limit2} we showed convergence of the rescaled walk $X_t/t$ in distribution.
The density function of the limit law had both the Dirac $\delta$-function and a continuous function with a compact support on an interval determined only by the coin operator $C$.
The limit theorem was linked to Theorem~\ref{th:limit1} via the quantity $\Delta$, which was both the coefficient of the Dirac $\delta$-function and the sum $\sum_{x\in\mathbb{Z}}\lim_{t\to\infty}\mathbb{P}(X_t=x)$.
When $\Delta=0$, localization doesn't occur from Theorem~\ref{th:limit1} and the limit distribution doesn't have the Dirac $\delta$-function from Theorem~\ref{th:limit2}. 

\begin{figure}[h]
\begin{center}
 \begin{minipage}{70mm}
  \begin{center}
  \includegraphics[scale=0.5]{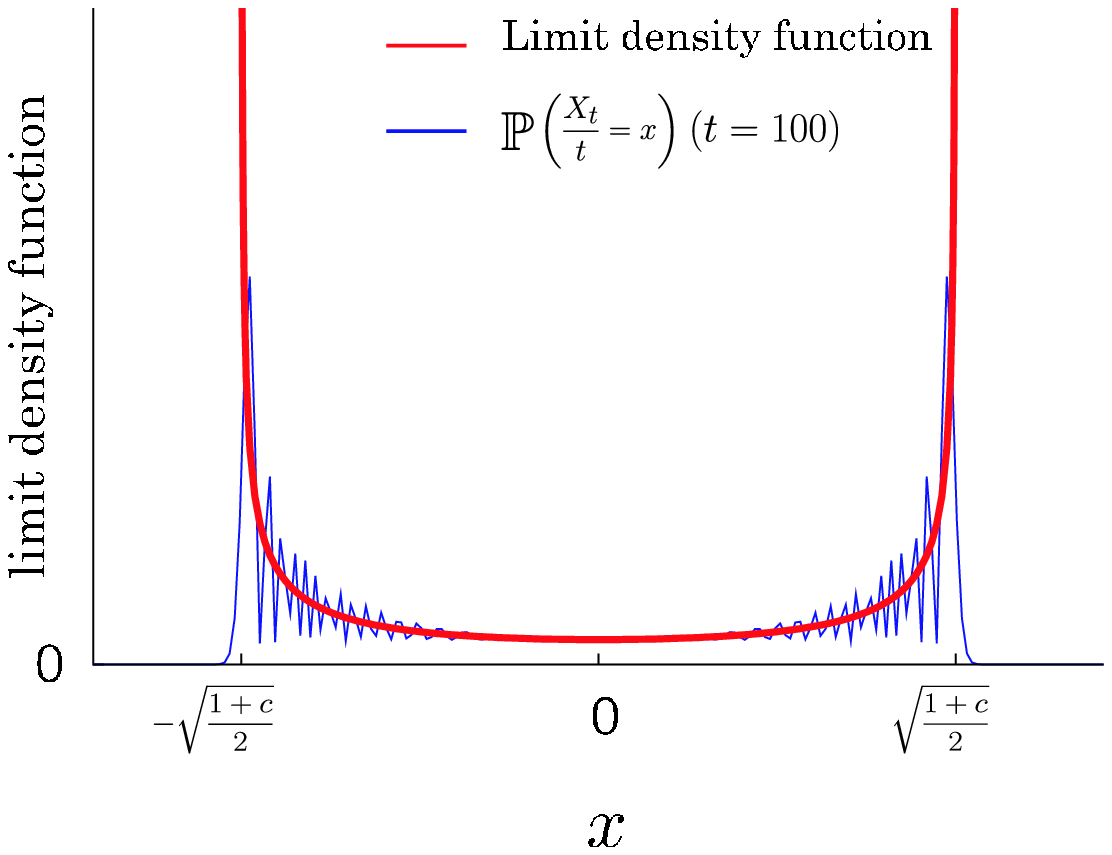}\\[2mm]
  (a) 3-state walk
  \end{center}
 \end{minipage}
 \begin{minipage}{70mm}
  \begin{center}
   \includegraphics[scale=0.5]{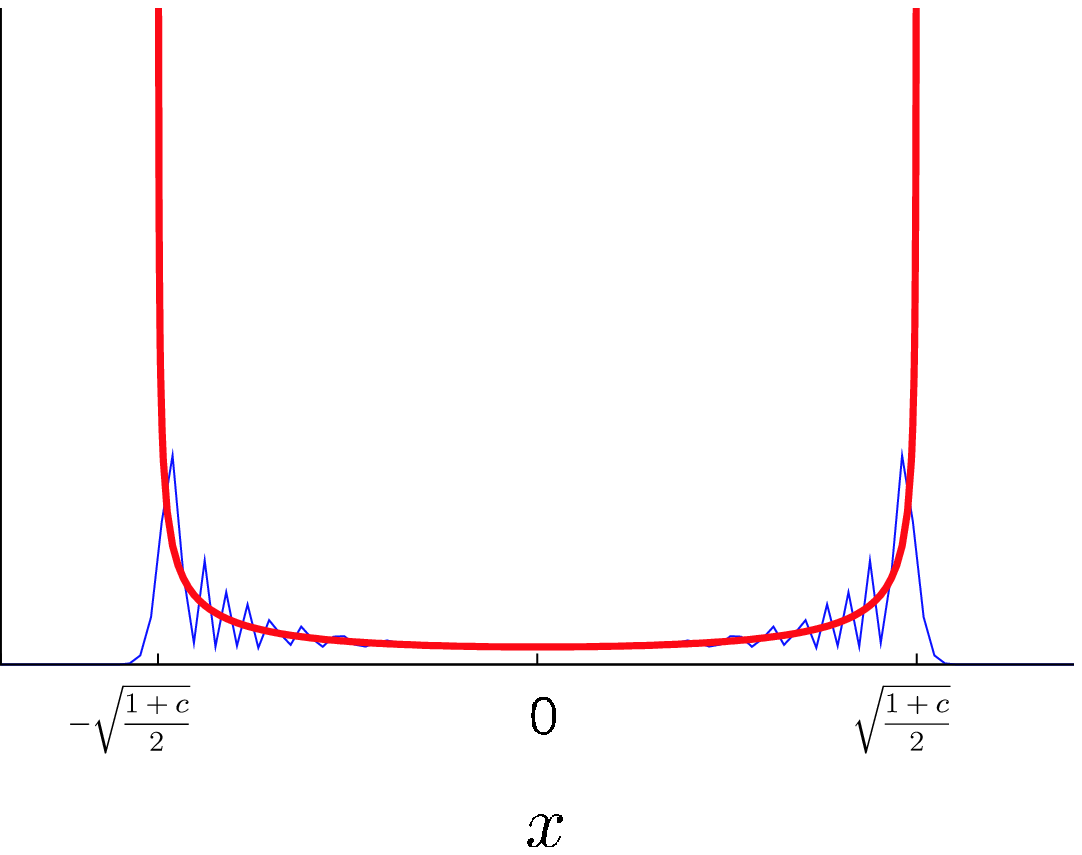}\\[2mm]
  (b) 2-state walk
  \end{center}
 \end{minipage}
\vspace{5mm}
\fcaption{Figure~(a) (resp. (b)) shows probability distribution at time $t=100$ and the limit density function of the 3-state (resp. 2-state) walk with $\alpha=\gamma=-\frac{1}{2},\beta=\frac{1}{\sqrt{2}}$ (resp. $\alpha=\frac{1}{\sqrt{2}},\beta=\frac{i}{\sqrt{2}}$) in the case of $\theta=\frac{\pi}{2}$.
The parameters $\alpha,\beta$ used in Fig.~(b) mean those in Konno~\cite{Konno2002,Konno2005}.
 }
\label{fig:2-state}
\end{center}
\end{figure}

Additionally, we can discuss a relation between the 3-state walk and a 2-state walk from the observation of Theorem~\ref{th:limit2}.
Under the condition $|\beta|=\sqrt{\frac{1-c}{2}}, \alpha=\gamma=-\frac{\sqrt{2}s}{2(1-c)}\beta$, the limit distribution of the 3-state walk does not have the Dirac $\delta$-measure and it has a continuous density
\begin{equation}
 \frac{\sqrt{1-\frac{1+c}{2}}}{\pi(1-x^2)\sqrt{\frac{1+c}{2}-x^2}}I_{(-\sqrt{\frac{1+c}{2}},\sqrt{\frac{1+c}{2}})}(x)\,dx.\label{eq:2-state}
\end{equation}
This is also a limit density of a 2-state walk (See Konno~\cite{Konno2002,Konno2005} and take $|a|=\sqrt{\frac{1+c}{2}}, \Re(a\alpha\overline{b\beta})=0$).
In more detail the limit distribution of the 3-state walk belongs to the class of the limit distribution of the 2-state walk if and only if $|\beta|=\sqrt{\frac{1-c}{2}}, \alpha=\gamma=-\frac{\sqrt{2}s}{2(1-c)}\beta$.
As is shown in Fig.~\ref{fig:2-state}, although the 3-state walk and the 2-state walk at any time are obviously distinct from each other, both take the same limit density function Eq.~(\ref{eq:2-state}) as $t\to\infty$.
Theorem~\ref{th:limit2} might be available to discuss the difference between the two walks from the viewpoint of physics.

\begin{figure}[h]
\begin{center}
 \begin{minipage}{50mm}
  \begin{center}
  \includegraphics[scale=0.4]{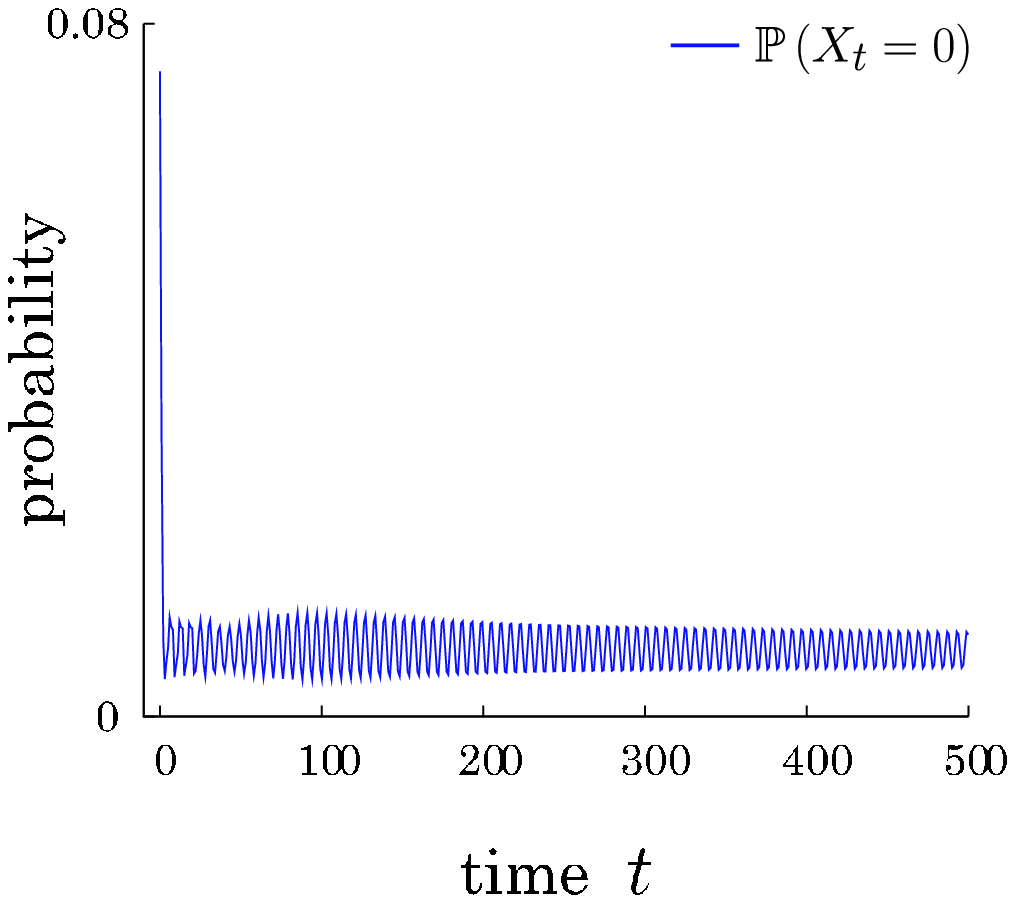}\\[2mm]
  (a) Example~\ref{ex:1}\\ $\alpha=\beta=0, \gamma=1$
  \end{center}
 \end{minipage}
 \begin{minipage}{40mm}
  \begin{center}
   \includegraphics[scale=0.4]{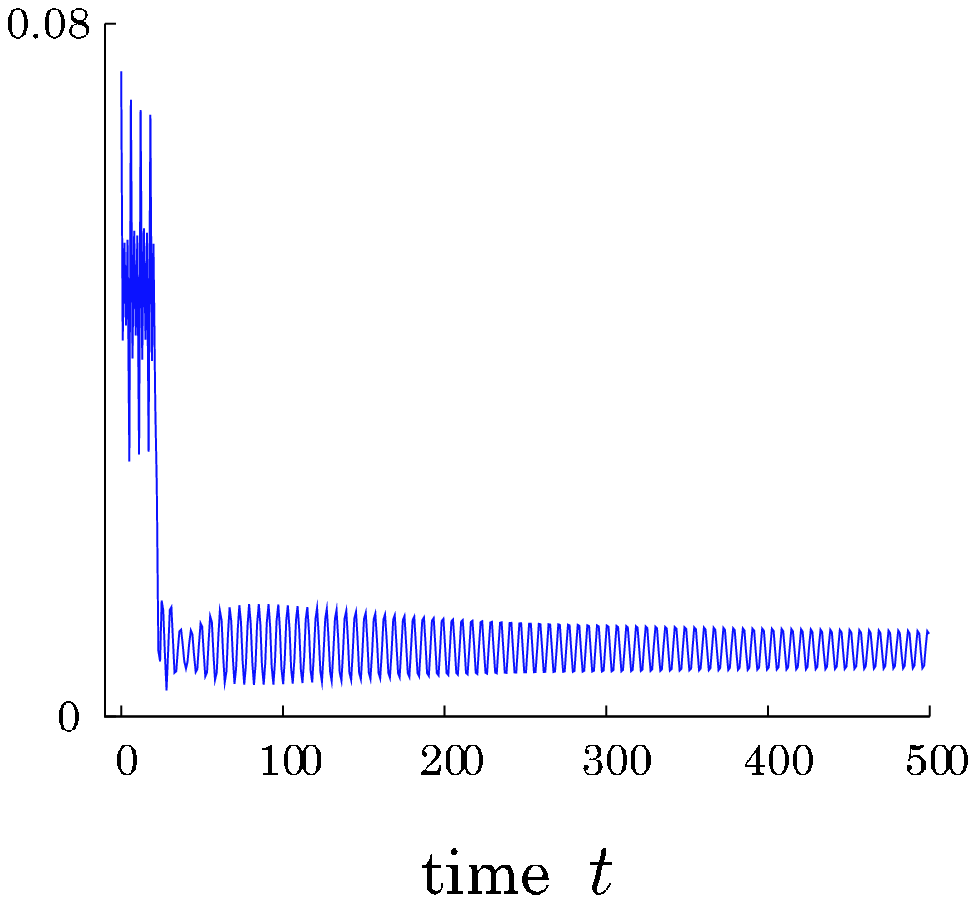}\\[2mm]
  (b) Example~\ref{ex:2}\\ $\alpha=1, \beta=\gamma=0$
  \end{center}
 \end{minipage}
 \begin{minipage}{40mm}
  \begin{center}
   \includegraphics[scale=0.4]{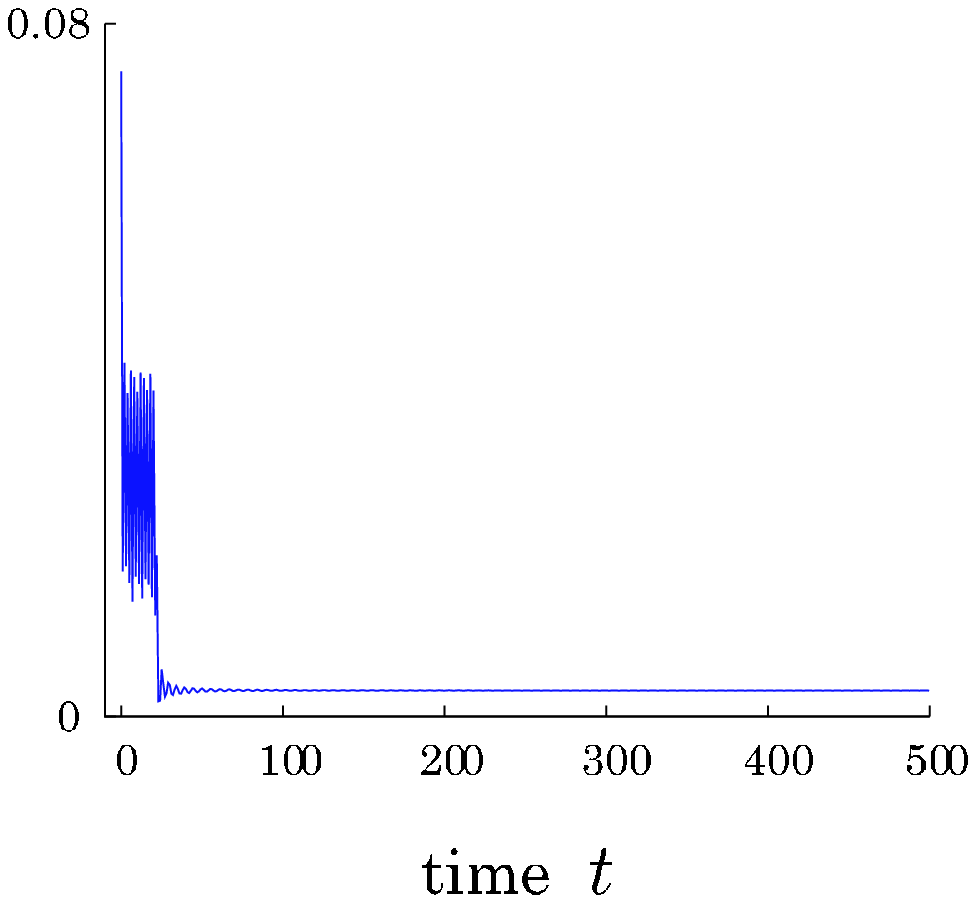}\\[2mm]
  (c) Example~\ref{ex:3}\\ $\alpha=-\gamma=\frac{1}{\sqrt{2}},\beta=0$
  \end{center}
 \end{minipage}
\end{center}
\vspace{5mm}
\fcaption{Figures~(a), (b) and (c) show the probabilities at the origin $\mathbb{P}(X_t=0)$ in Fig~\ref{fig:ex123} for time $t$, respectively.}
\label{fig:ex123_te}
\end{figure}

As an application of the 3-state walk, we created some uniform measures by using a delocalized initial state in Sec.~\ref{sec:application}.
It can be said that we have got a method to prepare a discrete uniform measure by using a quantum walk.
Equation~(\ref{eq:limit_measure_2}) was not probability distribution as well as the limit measure for the walk starting from the origin.
For example, since the limit value $\lim_{t\to\infty}\mathbb{P}(X_t=x)$ in Example~\ref{ex:1} is less than $\frac{1}{4n}$ for any $x\in\mathbb{Z}$, the sum over all positions does not exceed $\frac{1}{2}$.
One of the future problems for this application is to evaluate the convergence speed of the discrete measures for time $t$.
Seeing the time evolution of probability distribution in Figs.~\ref{fig:ex123}-(a) and -(b), we find those of Examples~\ref{ex:1} and \ref{ex:2} fluctuate around the limit values.
On the other hand, as you can see from Fig.~\ref{fig:ex123}-(c), it seems that probability distribution of Example~\ref{ex:3} converges to the limit earlier than the others.
In fact, Fig~\ref{fig:ex123_te} shows time-evolution of the probability at the origin $\mathbb{P}(X_t=0)$.
To applicate the uniform measures for uniform random values actually, we need to estate the difference between the limit values and probability distributions for time $t$ because we must take care of a realization of QWs within finite steps in a quantum system.

\nonumsection{Acknowledgements}
\noindent I am grateful to the Japan Society for the Promotion of Science for the support and to the Math. Dept. UC Berkeley for hospitality, and thank F. Alberto Gr{\"u}nbaum for useful comments.


\end{document}